\documentclass{article}
\usepackage{graphicx}
\usepackage{spconf}
\usepackage{color}
\usepackage{float}
\usepackage{amsmath,bm,amssymb,amsfonts,amsthm,epsfig}
\usepackage{graphicx}
\usepackage{psfrag}
\usepackage[caption=false]{subfig}
\usepackage{comment}
\usepackage{url}
\usepackage{cite} 

\graphicspath{{Figures/}}
\usepackage{epstopdf}
\usepackage{lipsum}

\newtheorem{definition}{Definition}

\newtheorem{proposition}{Proposition}
\newtheorem{thm}{Theorem}
\newtheorem*{thm*}{Theorem}

\newtheorem{remark}{Remark}
\newtheorem*{remark*}{Remark}

\usepackage[hidelinks]{hyperref}
\usepackage[capitalize]{cleveref}
\crefname{thm}{Theorem}{Theorems}

\newcounter{opteq}

\usepackage{arydshln}
\usepackage{mathrsfs}
\usepackage{bbm}

\usepackage{pifont}

\usepackage[toc,page]{appendix}

\newcommand{\txt}[1]{\text{\normalfont #1}}   

\newcommand{\zero}{\textcolor[rgb]{.7,.7,.7}{0}}

\DeclareMathOperator{\rank}{rank}
\DeclareMathOperator{\krank}{k-rank}

\DeclareMathOperator{\kron}{\otimes}
\DeclareMathOperator{\krao}{\odot}

\DeclareMathOperator{\tx}{t}
\DeclareMathOperator{\rx}{r}

\DeclareMathOperator{\ai}{{I}}
\DeclareMathOperator{\aii}{\txt{II}}

\newcommand{\Nt}{3}
\newcommand{\Nr}{4}
\newcommand{\NSigma}{8}
\newcommand{\figsize}{7}

\makeatletter
\newcounter{savesection}
\newcounter{apdxsection}
\renewcommand\appendix{\par
	\setcounter{savesection}{\value{section}}%
	\setcounter{section}{\value{apdxsection}}%
	\setcounter{subsection}{0}%
	\gdef\thesection{\@Alph\c@section}}
\newcommand\unappendix{\par
	\setcounter{apdxsection}{\value{section}}%
	\setcounter{section}{\value{savesection}}%
	\setcounter{subsection}{0}%
	\gdef\thesection{\@arabic\c@section}}
\makeatother

\usepackage{enumitem}
\usepackage{tikz}
\usetikzlibrary{tikzmark,calc,decorations.pathreplacing,shapes,backgrounds,patterns}
\usepackage{pgfplots} 
\pgfplotsset{compat=1.16}
\pgfplotsset{table/search path={Data},
        colormap={parula}{
            rgb255=(53,42,135)
            rgb255=(15,92,221)
            rgb255=(18,125,216)
            rgb255=(7,156,207)
            rgb255=(21,177,180)
            rgb255=(89,189,140)
            rgb255=(165,190,107)
            rgb255=(225,185,82)
            rgb255=(252,206,46)
            rgb255=(249,251,14)
        },
        every axis/.append style={
                    label style={font=\small},
                    tick label style={font=\small},  
                    title style={font=\small}, 
                    },
    contour/label node code/.code={%
        \node{$m=\pgfmathprintnumber{#1}$};}
    }
\usepgfplotslibrary{fillbetween}
\usetikzlibrary{pgfplots.groupplots}
\usepackage[percent]{overpic}

\crefname{enumi}{Array}{Arrays} 

\newcommand{\rd}[1]{\textcolor[rgb]{1,0,0}{#1}}
\newcommand{\bl}[1]{\textcolor[rgb]{0,0,0}{#1}}

\usepackage[hang,flushmargin]{footmisc} 

\newcommand\blfootnote[1]{%
  \begingroup
  \renewcommand\thefootnote{}\footnote{#1}%
  \addtocounter{footnote}{-1}%
  \endgroup
}

\makeatletter
\def\tikz@auto@anchor{%
    \pgfmathtruncatemacro\angle{atan2(\pgf@x,\pgf@y)-90}
    \edef\tikz@anchor{\angle}%
}
\makeatother

\title{Importance of array redundancy pattern in active sensing}
\name{\begin{tabular}{c} Robin Rajam\"{a}ki,
		~Piya Pal
		\end{tabular}}
\address{Department of Electrical and Computer Engineering, University of California San Diego, USA}

\begin{document}
	\setlength\belowcaptionskip{-15ex}
	%
	\maketitle
	\begin{abstract}		
    This paper further investigates the role of the array geometry and redundancy in active sensing. We are interested in the fundamental question of how many point scatterers can be identified (in the angular domain) by a given array geometry using a certain number of linearly independent transmit waveforms. We consider redundant array configurations (with repeated virtual transmit-receive sensors), which we have recently shown to be able to achieve their maximal identifiability while transmitting fewer independent waveforms than transmitters. Reducing waveform rank in this manner can be beneficial in various ways. For example, it may free up spatial resources for transmit beamforming. In this paper, we show that two array geometries with identical \bl{sum co-arrays}, and the same number of physical and virtual sensors, need not achieve equal identifiability---regardless of the choice of waveform of a fixed reduced rank. This surprising result establishes the important role the \emph{pattern} (not just the number) of repeated virtual sensors has in governing identifiability, and reveals the limits of compensating for unfavorable array geometries via waveform design.
    \blfootnote{This work was supported in part by grants ONR N00014-19-1-2256, ONR N00014-19-1-2227, NSF 2124929 and DE-SC0022165, as well as the Ulla Tuominen foundation and the Finnish Defence Research Agency.}
	\end{abstract}
	\setlength{\textfloatsep}{7pt}

	\section{Introduction}

\emph{Active} sensing and sparse arrays play a key role in numerous applications including autonomous sensing \cite{ma2020joint}, automotive radar \cite{sun2020mimoradar}, and emerging wireless systems for joint communications and sensing (JCS) \cite{liu2020joint,hassanien2016signaling}. The performance of such systems critically depends on the transmitted waveforms \cite{levanon2004radar,kumari2020adaptive,wang2020slowtime} as well as the employed transmit (Tx) and receive (Rx) array geometries \cite{wang2017coarrays,tohidi2019sparse,sedighi2021ontheperformance,sarangi2021beyond,sarangi2022single,sarangi2023superresolution}. In particular, multiple-input multiple-output (MIMO) systems can adjust the number of linearly independent waveforms or so-called \emph{waveform rank} (WR) to, e.g., trade off between Tx beamforming gain and field-of-view. In colocated MIMO radar \cite{bliss2003multiple}, full WR in the form of orthogonal waveforms is conventionally employed to maximize the number of identifiable scatterers. However, since \emph{identifiability} is upper bounded by the size of the \emph{\bl{sum co-array}} \cite{li2007onparameter}---consisting of the pairwise sums of the Tx-Rx sensor positions---a \emph{redundant} array, which has repeated virtual sensors, may actually achieve its maximal identifiability using a reduced WR \cite{rajamaki2023arrayinformed}. This enables redirecting spatial resources towards beamforming or, communications in the case of dual-function JCS systems. Additional advantages of redundant arrays include robustness to sensor failure \cite{liu2019robustnessi} and resilience to noise due to spatial averaging over repeated virtual sensors.

Until recently, relatively little was known about the impact of array redundancy and WR on identifiability. In our recent work, we showed that maximizing identifiability at a reduced WR requires the Tx waveform to be matched to the array geometry \cite{rajamaki2023arrayinformed}. Indeed, even two waveforms giving rise to \emph{identical} Tx beampatterns can yield different identifiability when employed by the same array. Hence, constraining the Tx beampattern without considering the joint Tx-Rx array geometry may result in suboptimal sensing performance. This perspective gives rise to yet unaddressed questions such as: Is proper waveform design \emph{sufficient} for maximizing identifiability? Can all array geometries of a given size and with identical \bl{sum co-arrays} achieve the same identifiability simply by choosing the waveforms (of fixed WR) suitably? 

This paper answers these question\bl{s} in the negative. Specifically, we demonstrate that two redundant array geometries can have the same number of physical sensors and identical uniform \bl{sum co-arrays}, yet different identifiability properties. Surprisingly, \emph{no} choice of waveforms can improve identifiability when employing an unfavorable redundant array geometry and a reduced WR. This novel insight reveals the impact that the configuration of redundant virtual sensors has on identifiability, and highlights the importance of judicious sparse array design, which is especially important in future resource-efficient active sensing systems, such as autonomous sensing and JCS.


\section{Signal model}
Let the support of unknown $K$-sparse vector $\bm{x}\in\mathbb{C}^V$ encode the angular directions of $K$ far field scatterers, which lie on a grid of $V\gg K$ candidate angles. For a single range-Doppler cell, the Rx vector of a colocated active sensing MIMO system in absence of noise can be modeled as \cite{bekkerman2006target,friedlander2012onsignal,rajamaki2023arrayinformed}%
\begin{align}
\bm{y}
=(\bm{S}\kron \bm{I})(\bm{A}_{\tx}\krao \bm{A}_{\rx})\bm{x}
=(\bm{S}\kron \bm{I})\bm{\Upsilon}\bm{A}\bm{x} \triangleq \bm{B}\bm{x}.
\label{eq:y_noiseless}
\end{align} 
Here, $\kron$ and $\krao$ denote the Kronecker and Khatri-Rao (columnwise Kronecker) products, respectively, and $\bm{I}$ is the $N_{\rx}\times N_{\rx}$ indentity matrix, where $N_{\rx}$ is the number of Rx sensors. Moreover, $\bm{S}\!\in\!\mathbb{C}^{T\times N_{\tx}}$ is a known deterministic \emph{spatio-temporal waveform matrix} (or space-time code \cite{jafarkhani2005spacetime}) whose columns represent signals of length $T$ launched by the $N_{\tx}$ Tx sensors. The effective Tx-Rx manifold matrix $ \bm{A}_{\tx}\krao\bm{A}_{\rx}\in\mathbb{C}^{N_{\tx}N_{\rx}\times V}$ models the phase shifts incurred by the narrowband radiation transmitted/received by the arrays. Since $\bm{A}_{\tx}\krao\bm{A}_{\rx}$ can have repeated rows, we may write $\bm{A}_{\tx}\krao\bm{A}_{\rx}\!=\!\bm{\Upsilon}\bm{A}$, where $\bm{A}\in\mathbb{C}^{N_\Sigma\times V}$ is the manifold matrix of a virtual array \bl{(independent of $\bm{S}$)} with $N_\Sigma\leq N_{\tx}N_{\rx}$ unique virtual sensors. This so-called \emph{sum co-array} \cite{hoctor1990theunifying}, denoted by $\mathbb{D}_\Sigma$, is defined as
\begin{align}
		\mathbb{D}_\Sigma  \triangleq \mathbb{D}_{\tx}+\mathbb{D}_{\rx}=\{d_{\tx} +d_{\rx}\ |\ d_{\tx}\in\mathbb{D}_{\tx}; d_{\rx}\in\mathbb{D}_{\rx}\},\label{eq:sca}
\end{align}
where $\mathbb{D}_{\tx}\triangleq\{d_{\tx}[n]\}_{n=1}^{N_{\tx}}$ and $\mathbb{D}_{\rx}\triangleq\{d_{\rx}[m]\}_{m=1}^{N_{\rx}}$ are the set of Tx and Rx sensor positions, respectively. 
 Furthermore, $\bm{\Upsilon}\in\{0,1\}^{N_{\tx}N_{\rx}\times N_\Sigma}$ is the so-called \emph{redundancy pattern} matrix mapping the $N_\Sigma\triangleq |\mathbb{D}_\Sigma|$ unique virtual sensors in $\mathbb{D}_\Sigma$ \bl{($|\mathbb{D}_\Sigma|$ denotes the cardinality of $\mathbb{D}_\Sigma$)} to the corresponding physical Tx-Rx sensor pairs $(d_{\tx},d_{\rx})$.
\begin{definition}[Redundancy pattern]
	The $ (n,\ell)$th entry of the binary redundancy pattern matrix $ \bm{\Upsilon}\in\{0,1\}^{N_{\tx}N_{\rx}\times N_\Sigma} $ is
	\begin{align*}
		\Upsilon_{n,\ell}\!\triangleq\!
  \bl{\begin{cases}
      1,&\txt{if } d_{\tx}\big[\lceil \frac{n}{N_{\rx}} \rceil\big]\!+\!d_{\rx}\big[n\!-\!(\lceil\!\frac{n}{N_{\rx}}\rceil\!-\!1) N_{\rx}\big]\!=\!d_\Sigma[\ell]\\
      0,& \txt{otherwise}.
  \end{cases}}
	\end{align*}
	Here, $d_{\Sigma}[\ell]\in\mathbb{D}_{\Sigma}$ is the $\ell$th sum co-array element position\bl{, and $\lceil\cdot\rceil$ denotes the ceiling function}.
\end{definition}
An array is \emph{redundant} if $N_\Sigma<N_{\tx}N_{\rx}$ and nonredundant if $N_\Sigma=N_{\tx}N_{\rx}$. Furthermore, sum co-array $\mathbb{D}_\Sigma$ is \emph{contiguous} if $ \mathbb{D}_\Sigma\!=\!\{0,1,\ldots,N_\Sigma-1\}$, where $\mathbb{D}_{\tx},\mathbb{D}_{\rx}\supseteq\{0\}$ are assumed to be non-negative integer sets describing the normalized Tx/Rx sensor positions in units of half \bl{the} carrier wavelength. \bl{For a contiguous sum co-array, the $(\ell,i)$th entry of $\bm{A}$ is thus $A_{\ell,i}=\exp(j\pi (\ell-1)\sin\theta_i)$, where $\theta_i\in[-\tfrac{\pi}{2},\tfrac{\pi}{2})$.}

We are interested in understanding the interplay between the spatial sensing geometry (captured by $\bm{\Upsilon}\bm{A}$) and the waveform $\bm{S}$ that maximizes the number of identifiable scatterers (i.e., the size of the support of $\bm{x}$). It is well-known that the sparsest solution to $\bm{y}=\bm{B}\bm{x}$ is unique if and only if $K\leq\frac{1}{2}\krank(\bm{B})$ \cite{donoho2003optimally,pal2015pushing}, where $\krank(\bm{B})$ denotes the Kruskal rank\footnote{The Kruskal rank of matrix $\bm{B}$, denoted $\krank(\bm{B})$, is the largest integer $r$ such that \emph{every} $r$ columns of $\bm{B}$ are linearly independent.} of matrix $\bm{B}$. Due to the spatio-temporal structure of $\bm{B}$, its Kruskal rank is determined by the (i) sum co-array, which is modeled by virtual manifold matrix $\bm{A}$, (ii) redundancy pattern $\bm{\Upsilon}$, and (iii) waveform matrix $\bm{S}$. A key quantity of interest is the waveform rank (WR), defined as
\begin{align}
	N_s\triangleq \rank(\bm{S}). \label{eq:rank_S}
\end{align}
The case $N_s=1$ corresponds to the phased array, whereas a canonical example of $N_s=N_{\tx}$ is (orthogonal) MIMO radar, which is known to achieve maximal Kruskal rank $\krank(\bm{B})=N_\Sigma$ \cite{li2007onparameter,rajamaki2023arrayinformed}. Advantages of a reduced waveform rank $N_s<N_{\tx}$ include improved beamforming gain on transmit, fewer costly RF chains at the transmitter, and possibly decreased transmission time, since $T\geq N_s$. It is therefore important to understand (a) if maximal Kruskal rank can be attained for a given array at a reduced $N_s$, and (b) which choices of $\bm{S}$ enable this. \bl{To answer these questions}, we briefly review the key ideas of \emph{array-informed waveform design} introduced in \cite{rajamaki2023arrayinformed}.

    \section{Array-informed waveform design in a nutshell}\label{sec:review}
    
\begin{figure}
    \centering
    		\begin{tikzpicture}
			\begin{axis}[width=1*\figsize cm,height=1*\figsize cm,yticklabel shift = 0 pt,ylabel shift=-10 pt,xticklabel shift = 0 pt,xlabel shift = 0 pt,xmin=1,xmax=\Nt,ymin=1,ymax=\Nt*\Nr,
				xlabel={Rank of waveform matrix, $N_s=\rank(\bm{S})$},xtick={1,ceil(\NSigma/\Nr),\Nt},
				xticklabels={1,$\lceil N_\Sigma/N_{\rx}\rceil$,$N_{\tx}$},xtick pos=bottom,
				ylabel={Kruskal rank of sensing matrix, $\krank(\bm{B})$}, ytick={1,\Nr},yticklabels={1,$N_{\rx}$},
				extra y ticks={\NSigma,\Nt*\Nr},
				extra y tick labels={$N_\Sigma$,$N_{\tx}N_{\rx}$},
				extra y tick style ={yticklabel pos=right},
				legend style = {
					at={(0.47,0.73)},anchor = south,legend columns=1,draw=none,fill=none},legend cell align={left}]
				\addplot[name path=f,domain=1:\Nt,samples=\Nt, thick,black]{min(x*\Nr,\NSigma)};
				\addlegendentry{$\min(N_sN_{\rx},N_\Sigma)$}
				
				\path[name path=axis] (axis cs:1,1) -- (axis cs:\Nt,1);
				\addplot
				[color=lightgray,fill=lightgray]
				fill between[of=f and axis];
				\addlegendentry{Design space}
				
                \addplot[name path=f2,domain=ceil(\NSigma/\Nr):\Nt,samples=3,thick,red,dotted,forget plot]{\NSigma};
				\path[name path=axis2] (axis cs:{ceil(\NSigma/\Nr)},1) -- (axis cs:\Nt,1);
				\addplot
				[pattern =north east lines, pattern color = red]
				fill between[of=f2 and axis2];			
				\addlegendentry{Redundancy-limited regime}
                \addplot[only marks,mark=triangle*,mark size=3] coordinates {
            					({ceil(\NSigma/\Nr)},\NSigma)
				};                
                \node[shift={(0cm,0cm)},pin={[pin distance=3.5cm]234:\footnotesize{\shortstack{Optimal point\\ (redundancy-limited\\ min. waveform rank)}}}] at (axis cs:({ceil(\NSigma/\Nr)},\NSigma) {};
                
			\end{axis}
		\end{tikzpicture}
    \caption{Array-dependent trade-off between $N_s$ and $\krank(\bm{B})$. The set of optimal operating points---in terms of identifiability---are given by the maximal Kruskal rank in \eqref{eq:max_krank}. Optimal point $(\big\lceil \frac{N_\Sigma}{N_{\rx}}\big\rceil,N_\Sigma)$ represents the minimum waveform rank achieving the maximum (redundancy-limited) Kruskal rank. Here, $N_{\tx}$, $N_{\rx}$ and $N_\Sigma$ denote the number of Tx, Rx, and virtual sensors, respectively.}
    \label{fig:design_space}
\end{figure}
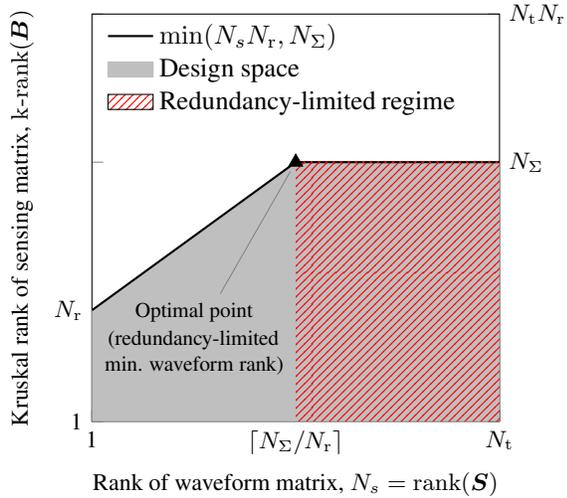

    
	\cref{fig:design_space} illustrates the range of values that the Kruskal rank of $\bm{B}$ may assume as a function of the WR given an arbitrary array with $N_{\tx}$ Tx sensors, $N_{\rx}$ Rx sensors, and $N_\Sigma$ virtual sensors \cite{rajamaki2023arrayinformed}.\footnote{There can be multiple arrays with the same $N_\Sigma$ but different redundancy patterns $\bm{\Upsilon}$ for a given $N_{\tx},N_{\rx}$.} The design space (shaded area) is upper bounded by the \emph{maximal} Kruskal rank \cite{rajamaki2023arrayinformed}
    \begin{align}
	    \krank(\bm{B})\leq \min(N_sN_{\rx},N_\Sigma),\label{eq:max_krank}
	\end{align}
    which is a piecewise linear function in $N_s$ tracing the set of identifiability-maximizing operating points. \cref{eq:max_krank} reveals the existence of an \emph{optimal} operating point $N_s=\lceil N_\Sigma/N_{\rx} \rceil $, which is the minimum WR needed to attain maximal Kruskal rank $N_\Sigma$. A key observation is that setting $N_s> \lceil N_\Sigma/N_{\rx} \rceil$ can be wasteful for redundant arrays, since $N_{\tx}-\lceil N_\Sigma/N_{\rx}\rceil $ Tx degrees of freedom could instead be used to, e.g., beamform, or serve users in dual-function JCS systems, without sacrificing identifiability.
 
The upper bound in \eqref{eq:max_krank} can be attained for any $N_{\tx},N_{\rx}, N_s$ and suitable values of $N_\Sigma$ \cite{rajamaki2023arrayinformed}. This requires both proper array design and ``matching'' waveform matrix $\bm{S}$ to the array geometry or redundancy pattern $\bm{\Upsilon}$.
We call this ``array-informed waveform design''. Intuitively, an $\bm{S}$ that maximizes identifability should minimize the dimension of the intersection between the null space of $\bm{S}\kron\bm{I}$ and range space of $\bm{\Upsilon}$. In the redundancy-limited regime $N_s\geq N_\Sigma/N_{\rx} $, where Kruskal rank $N_\Sigma$ could potentially be attained, the following necessary and sufficient condition for $\krank(\bm{B})=N_\Sigma$ holds.
\begin{proposition}[Redundancy-limited waveform rank {\cite[Theorem~2]{rajamaki2023arrayinformed}}]\label{thm:gen_restr_iff}
If $\rank(\bm{S})\geq N_\Sigma/N_{\rx}$, then
\begin{align}
    \krank(\bm{B})=N_\Sigma \iff 
    \begin{cases}
        \krank(\bm{A})=N_\Sigma \txt{ and}\\
        \rank((\bm{S}\kron\bm{I})\bm{\Upsilon})=N_\Sigma.
    \end{cases}\label{eq:gen_restr_iff}
\end{align}
\end{proposition}
The utility of \cref{thm:gen_restr_iff} stems from the fact that if the sum co-array is contiguous, then $\bm{A}$ is a Vandermonde matrix and $\krank(\bm{A})=N_\Sigma$. Hence, verifying the computationally challenging Kruskal rank condition $\krank(\bm{B})\!=\!N_\Sigma$ reduces to the simpler condition $\rank((\bm{S}\kron\bm{I})\bm{\Upsilon})\!=\!N_\Sigma$. 

This paper delves deeper into the question ``what combination of array geometry and waveform can achieve maximal Kruskal rank in the redundancy-limited regime?'' When $N_s=N_{\tx}$, it can easily be verified that any array with a contiguous sum co-array achieves $\krank(\bm{B})=N_\Sigma$, regardless of the choice of $\bm{S}$ (of rank $N_{\tx}$) \cite{rajamaki2023arrayinformed}. However, the answer is not so obvious when $N_s<N_{\tx}$. The optimal point $N_s=\lceil N_\Sigma/N_{\rx} \rceil $ in \cref{fig:design_space} is of particular interest. Hence, the remainder of the paper focuses on \emph{whether Kruskal rank $N_\Sigma$ is always attainable with minimum WR}. Specifically, given \emph{any} array geometry with a contiguous co-array of a fixed size $N_\Sigma$, does there always exists a choice of $\bm{S}$ such that $\krank(\bm{B})=N_\Sigma$ when $N_s=\lceil N_\Sigma/N_{\rx}\rceil$? Interestingly, the answer is \emph{no}, as we show next.
     
\section{Importance of redundancy pattern for maximizing identifiability}\label{sec:main}
 This section demonstrates that the maximum \emph{achievable} Kruskal rank of a given array geometry depends on redundancy pattern $\bm{\Upsilon}$; \emph{not} solely on tuple $(N_{\tx},N_{\rx}, N_\Sigma, N_s)$. Hence, the solid black line in \cref{fig:design_space} may be attained for all $N_s$ by one array geometry, but not another with the same contiguous co-array, and number of physical/virtual sensors $N_{\tx},N_{\rx},N_\Sigma$---regardless of the choice of waveform matrix $\bm{S}$.
    \begin{thm}\label{thm:array_bad_good}
 There exists two array geometries with contiguous sum co-arrays and the same $N_{\tx},N_{\rx},N_\Sigma$ such that the associated spatio-temporal sensing matrices $\bm{B}_{\ai}$ and $\bm{B}_{\aii}$ obey $\krank(\bm{B}_{\ai})\!<\!\!N_\Sigma$ and $\krank(\bm{B}_{\aii})\!=\!N_\Sigma$ when $N_s=\lceil N_\Sigma/N_{\rx}\rceil$.
\end{thm}
\begin{remark}
    \cref{thm:array_bad_good} shows the importance of proper array design for maximizing identifiability, as two arrays with the same (contiguous) sum co-array and number of physical/virtual sensors need not achieve equal identifiablity. An unfavorable redundancy pattern can thus limit identifiability.
\end{remark}
To show existence in \cref{thm:array_bad_good}, \bl{it suffices to} construct two array configurations for specific values of $N_{\tx},N_{\rx}$, and $N_\Sigma $. This conveys the essential idea of the proof technique. A more general proof, including extensions to other values of $(N_{\tx},N_{\rx},N_\Sigma,N_s)$, is part of ongoing work.

    \subsection{Proof sketch of \cref{thm:array_bad_good}}\label{sec:proof}
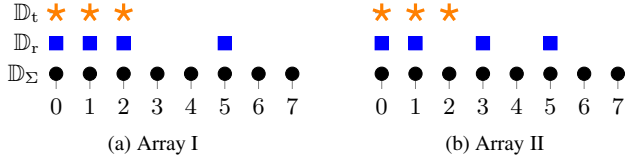
\begin{figure}
     	\renewcommand{\NSigma}{8} 
     	\centering
     	\subfloat[\cref{i:array_bad}]{  
     		\begin{tikzpicture} 
     			\begin{axis}[width=0.7*\figsize cm,height=\figsize*0.4 cm,ytick={-1,0,1},yticklabels={$\mathbb{D}_{\Sigma}$,$\mathbb{D}_{\rx}$,$\mathbb{D}_{\tx}$},xmin=-0.2,xmax=\NSigma-1+0.2,ymin=-1.5,ymax=1.5,xtick={0,1,...,\NSigma-1},title style={yshift=-7 pt},xticklabel shift = 0 pt,xtick pos=bottom,ytick pos=left,axis line style={draw=none}]
     				\addplot[orange,only marks,mark=star,mark size=3.5,very thick] coordinates {
     					(0,1)
     					(1,1)
     					(2,1)};
     				\addplot[blue,only marks,mark=square*,mark size=2.5] coordinates {
     					(0,0)
     					(1,0)
     					(2,0)
     					(5,0)};
     				\addplot[only marks,mark=*,mark size=2.5] expression [domain=0:(\NSigma-1), samples=\NSigma] {-1};
     			\end{axis}
     		\end{tikzpicture}\label{fig:array_bad}
     	}\hfill
     	\subfloat[\cref{i:array_good}]{
     		\begin{tikzpicture} 
     			\begin{axis}[width=0.7*\figsize cm,height=\figsize*0.4 cm,ytick={-1,0,1},
     				yticklabels={},
     				xmin=-0.2,xmax=\NSigma-1+0.2,ymin=-1.5,ymax=1.5,xtick={0,1,...,\NSigma-1},title style={yshift=-7 pt},xticklabel shift = 0 pt,xtick pos=bottom,ytick pos=left,axis line style={draw=none}]
     				\addplot[orange,only marks,mark=star,mark size=3.5,very thick] coordinates {
     					(0,1)
     					(1,1)
     					(2,1)};
     				\addplot[blue,only marks,mark=square*,mark size=2.5] coordinates {
     					(0,0)
     					(1,0)
     					(3,0)
     					(5,0)};
     				\addplot[only marks,mark=*,mark size=2.5] expression [domain=0:(\NSigma-1), samples=\NSigma] {-1};
     			\end{axis}
     		\end{tikzpicture}\label{fig:array_good}
     	}
     	\caption{Array geometries achieving different identifiability despite having an equal number of physical sensors and identical contiguous sum co-arrays. \cref{i:array_good} attains the maximal Kruskal rank $N_\Sigma\!=\!8$ when $N_s\!=\!N_\Sigma/N_{\rx}\!=\!2$, but not \cref{i:array_bad}.}
     	\label{fig:arrays}
     \end{figure}

      Consider the following two array geometries:
     \begin{enumerate}[label=\Roman*.,ref=\Roman*]
     	\item $ \mathbb{D}_{\tx}=\{0,1,2\}$ and $\mathbb{D}_{\rx}=\{0,1,\rd{2},5\}$ \label{i:array_bad}
     	\item $ \mathbb{D}_{\tx}=\{0,1,2\}$ and $\mathbb{D}_{\rx}=\{0,1,\rd{3},5\}$. \label{i:array_good}
     \end{enumerate}
     These configurations differ only in the position of a single Rx sensor (highlighted in red), as illustrated in \cref{fig:arrays}. The corresponding redundancy patterns $\bm{\Upsilon}_{\ai}$ and $\bm{\Upsilon}_{\aii}$ are
\begin{align*}
\begingroup 
\setlength\arraycolsep{3pt}
    \bm{\Upsilon}_{\ai}=
    \underset{\txt{\small \cref{i:array_bad}}}{
    \begin{bmatrix}
1 & \zero & \zero & \zero & \zero & \zero & \zero & \zero 
\\
 \zero & 1 & \zero & \zero & \zero & \zero & \zero & \zero 
\\
 \zero & \zero & \rd{1} & \zero & \zero & \zero & \zero & \zero 
\\
 \zero & \zero & \zero & \zero & \zero & 1 & \zero & \zero 
\\
 \zero & 1 & \zero & \zero & \zero & \zero & \zero & \zero 
\\
 \zero & \zero & 1 & \zero & \zero & \zero & \zero & \zero 
\\
 \zero & \zero & \zero & \rd{1} & \zero & \zero & \zero & \zero 
\\
 \zero & \zero & \zero & \zero & \zero & \zero & 1 & \zero 
\\
 \zero & \zero & 1 & \zero & \zero & \zero & \zero & \zero 
\\
 \zero & \zero & \zero & 1 & \zero & \zero & \zero & \zero 
\\
 \zero & \zero & \zero & \zero & \rd{1} & \zero & \zero & \zero 
\\
 \zero & \zero & \zero & \zero & \zero & \zero & \zero & 1 
\end{bmatrix}},
\bm{\Upsilon}_{\aii}=
\underset{\txt{\small \cref{i:array_good}}}{
\begin{bmatrix}
1 & \zero & \zero & \zero & \zero & \zero & \zero & \zero 
\\
 \zero & 1 & \zero & \zero & \zero & \zero & \zero & \zero 
\\
 \zero & \zero & \zero & \rd{1} & \zero & \zero & \zero & \zero 
\\
 \zero & \zero & \zero & \zero & \zero & 1 & \zero & \zero 
\\
 \zero & 1 & \zero & \zero & \zero & \zero & \zero & \zero 
\\
 \zero & \zero & 1 & \zero & \zero & \zero & \zero & \zero 
\\
 \zero & \zero & \zero & \zero & \rd{1} & \zero & \zero & \zero 
\\
 \zero & \zero & \zero & \zero & \zero & \zero & 1 & \zero 
\\
 \zero & \zero & 1 & \zero & \zero & \zero & \zero & \zero 
\\
 \zero & \zero & \zero & 1 & \zero & \zero & \zero & \zero 
\\
 \zero & \zero & \zero & \zero & \zero & \rd{1} & \zero & \zero 
\\
 \zero & \zero & \zero & \zero & \zero & \zero & \zero & 1
\end{bmatrix}}.
\endgroup
\end{align*}
Both arrays have a contiguous sum co-array with $N_\Sigma=8$ virtual sensors. Nevertheless, \cref{i:array_bad} \emph{cannot} achieve maximal Kruskal rank for $N_s=N_\Sigma/N_{\rx}=2$, unlike \cref{i:array_good}. This is shown next using \cref{thm:gen_restr_iff}, which reduces to analyzing the rank of $\bm{W}_i\triangleq(\bm{S}\kron\bm{I})\bm{\Upsilon}_i$, $i\in\{\ai,\aii\}$, since $\bm{A}$ is Vandermonde and has full Kruskal rank both for \cref{i:array_bad,i:array_good}.
     
By assumption, $N_s=2$. Let also $T=2$ such that
\begin{align}
    \bm{S}=
    \begin{bmatrix}
        s_{11}&s_{12}&s_{13}\\
        s_{21}&s_{22}&s_{23}
    \end{bmatrix}.\label{eq:S_rank_2}
\end{align}
We may consider \eqref{eq:S_rank_2} without loss of generality for the purpose of applying \cref{thm:gen_restr_iff}, since for any $\bm{S}'\!\in\!\mathbb{C}^{T'\times N_{\tx}}$ such that $\rank(\bm{S}')\!=\!N_s$, any rank-revealing decomposition $\bm{S}'\!=\!\bm{U}\bm{S}$, where $\bm{U}\!\in\!\mathbb{C}^{T'\times N_s}$ and $\bm{S}\in\mathbb{C}^{N_s\times N_{\tx}}$ have full column and row rank, respectively (the columns of $\bm{U}$ span the range space of $\bm{S}'$), implies $\rank((\bm{S}'\kron\bm{I})\bm{\Upsilon})\!=\!\rank((\bm{U}\kron\bm{I})(\bm{S}\kron\bm{I})\bm{\Upsilon})\!=\!\rank((\bm{S}\kron\bm{I})\bm{\Upsilon})$.

In case of \cref{i:array_bad}, $\bm{W}_{\ai}$ reduces to
\begin{align*}
\bm{W}_{\ai}=
\begin{bmatrix}
s_{11} & s_{12} & s_{13} & \zero & \zero & \zero & \zero & \zero 
\\
 \zero & s_{11} & s_{12} & s_{13} & \zero & \zero & \zero & \zero 
\\
 \zero & \zero & s_{11} & s_{12} & s_{13} & \zero & \zero & \zero 
\\
 \zero & \zero & \zero & \zero & \zero & s_{11} & s_{12} & s_{13} 
\\
 s_{21} & s_{22} & s_{23} & \zero & \zero & \zero & \zero & \zero 
\\
 \zero & s_{21} & s_{22} & s_{23} & \zero & \zero & \zero & \zero 
\\
 \zero & \zero & s_{21} & s_{22} & s_{23} & \zero & \zero & \zero 
\\
 \zero & \zero & \zero & \zero & \zero & s_{21} & s_{22} & s_{23} 
\end{bmatrix}.
\end{align*}
The last three columns of $\bm{W}_{\ai}$ are clearly linearly dependent. Hence, $\bm{W}_{\ai}$ is rank-deficient when $N_s=2$, i.e., $\rank(\bm{W}_{\ai})\!<\!N_\Sigma\!\implies\!\krank(\bm{B}_{\ai})\!<\!N_\Sigma$, where $\bm{B}_{\ai}\!\triangleq\!\bm{W}_{\ai}\bm{A}$.

In case of \cref{i:array_good}, $\bm{W}_{\aii}$ evaluates to
\begin{align*}
\bm{W}_{\aii}=
\begin{bmatrix}
s_{11} & s_{12} & s_{13} & \zero & \zero & \zero & \zero & \zero 
\\
 \zero & s_{11} & s_{12} & s_{13} & \zero & \zero & \zero & \zero 
\\
 \zero & \zero & \zero & s_{11} & s_{12} & s_{13} & \zero & \zero 
\\
 \zero & \zero & \zero & \zero & \zero & s_{11} & s_{12} & s_{13} 
\\
 s_{21} & s_{22} & s_{23} & \zero & \zero & \zero & \zero & \zero 
\\
 \zero & s_{21} & s_{22} & s_{23} & \zero & \zero & \zero & \zero 
\\
 \zero & \zero & \zero & s_{21} & s_{22} & s_{23} & \zero & \zero 
\\
 \zero & \zero & \zero & \zero & \zero & s_{21} & s_{22} & s_{23}
\end{bmatrix}.
\end{align*}
There exists infinitely many choices of $\bm{S}$ yielding a full rank $\bm{W}_{\aii}$. For example, setting $s_{11},s_{13},s_{22}\in\mathbb{C}\setminus\{0\}$ and $s_{12}=s_{21}=s_{23}=0$ can be verified to ensure that $\rank(\bm{S})\!=\!2$ and $\rank(\bm{W}_{\aii})\!=\!N_\Sigma\!\implies\!\krank(\bm{B}_{\aii})\!=\!N_\Sigma$ (since $\bm{W}_{\aii}$ has full column rank), where $\bm{B}_{\aii}\!\triangleq\!\bm{W}_{\aii}\bm{A}$.\hfill$\blacksquare$\vspace{-0.2cm}

\section{Numerical example}
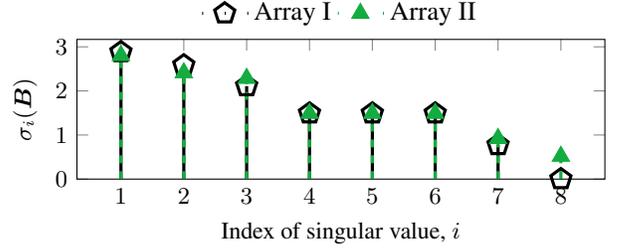
\begin{figure}
	\centering
	\begin{tikzpicture} 
		\begin{axis}[width=1\linewidth,height=0.4\linewidth,ylabel={$\sigma_i(\bm{B})$},xlabel= {Index of singular value, $i$},ymin=0,legend style = {at={(0.5,1.03)},anchor=south,draw=none,fill=none},legend columns=4,xtick={1,2,...,8}]
			\addplot+[ycomb,black, mark = pentagon,mark options={scale=2}, very thick] table[x=n,y =sigma]{Data/t_I_sv_B.dat};
			\addlegendentry{\cref{i:array_bad}}
			\addplot+[ycomb,red!10!green!75!blue, very thick,dashed, mark = triangle*,mark options={scale=1.5,solid}] table[x=n,y =sigma]{Data/t_II_sv_B.dat};%
			\addlegendentry{\cref{i:array_good}}
		\end{axis}%
	\end{tikzpicture}\vspace{-0.5cm}
\caption{Singular values of sensing matrix $\bm{B}$. This is rank-deficient in case of \cref{i:array_bad}---implying reduced Kruskal rank.}\label{fig:B_sv}
\end{figure}

We illustrate the implication of \cref{thm:array_bad_good} through a numerical example. \cref{fig:B_sv} shows the singular values of $\bm{B}$ for \cref{i:array_bad,i:array_good} given rank-2 waveform matrix $\bm{S}$ in \eqref{eq:S_rank_2} with $s_{11}=s_{22}=s_{13}=1/\sqrt{3}$ and $s_{12}=s_{21}=s_{23}=0$. The smallest singular value of \cref{i:array_bad} is zero, whereas that of \cref{i:array_good} is nonzero. Hence the Kruskal rank of \cref{i:array_bad} is no larger than $N_\Sigma-1=7$, since $\krank(\bm{B})\leq \rank(\bm{B})$. \cref{i:array_bad} can thus unambiguously identify at most $ \lfloor 7/2\rfloor=3$ scatterers. In contrast, \cref{i:array_good} can identify $4$ scatterers since its Kruskal rank is $N_\Sigma=8$. This is demonstrated in \cref{fig:raw_noiseless}, which shows a configuration of $K=4$ scatterers (grid size $V=16$) identifiable by \cref{i:array_good} but not \cref{i:array_bad}. This example is representative of an autonomous sensing scenario where identifying weak scatterers (pedestrians) close to strong ones (vehicles) is critical. We identified the scatterers by solving the following optimization problem by exhaustive search: $\txt{minimize}_{\bm{z}\in\mathbb{C}^V}\ \|\bm{z}\|_0$ subject to $\bm{y}_i=\bm{B}_i\bm{z}$, where $i\in\{\ai,\aii\}$ refers to \cref{i:array_bad,i:array_good}, respectively. We note that infinitely many unidentifiable scatterer configurations, such as the one in \cref{fig:raw_noiseless}, can be generated by straightforward linear algebraic manipulations when $\krank(\bm{B})<2K$.

\begin{figure}
	\begin{tikzpicture}
		\begin{groupplot}[
			group style={
				group size=2 by 1,
				x descriptions at=edge bottom,
				horizontal sep=0.7cm,
				vertical sep=1cm,
			},
			footnotesize,
			width=4.9cm,
			height=3.2cm,
			xlabel={Angle, $\theta$},
			ylabel=Magnitude,
			ymin=0,ymax=1,xmin=-pi/2,xmax=pi/2,xlabel={Angle, $\theta$},xtick={-pi/2,-pi/4,0,pi/4,pi/2},xticklabels={$-\frac{\pi}{2}$,$-\frac{\pi}{4}$,,$\frac{\pi}{4}$,$\frac{\pi}{2}$},xlabel shift = {-15 pt},ymode=linear
			]
			\nextgroupplot[%
			,legend to name=grouplegend
			,mark=none,legend style = {draw=none,fill=none},legend columns=2,
			title={(a) \cref{i:array_bad}},title style={at={(0.5,-0.7)}}
			]
			\addplot+[ycomb,black, mark = square,mark options={scale=1}] table[x=theta,y =sig]{Data/gt.dat};     
			\addlegendentry{Ground truth, $\bm{x}$}
			\addplot+[ycomb,red,mark=*,mark options={red,scale=.6,solid}, thick,dashed] table[x=theta,y=sig]{Data/t_I_2e-16.dat};
			\addlegendentry{Estimate, $\bm{z}$}
			\nextgroupplot[title={(b) \cref{i:array_good}},title style={at={(0.5,-0.7)}},ylabel={},yticklabels={}]
			\addplot+[ycomb,black, mark = square,mark options={scale=1}] table[x=theta,y =sig]{Data/gt.dat};     
			\addplot+[ycomb,red,mark=*,mark options={red,scale=.6,solid}, thick,dashed] table[x=theta,y=sig]{Data/t_II_2e-16.dat};
		\end{groupplot}
		\node at (group c1r1.north) [anchor=north, yshift=0.7cm, xshift=1.7cm] {\ref{grouplegend}};
	\end{tikzpicture}\vspace{-0.5cm}
	\caption{Scatterer configuration that cannot be identified by \cref{i:array_bad}, but can be identified by \cref{i:array_good}.}\label{fig:raw_noiseless}
\end{figure}
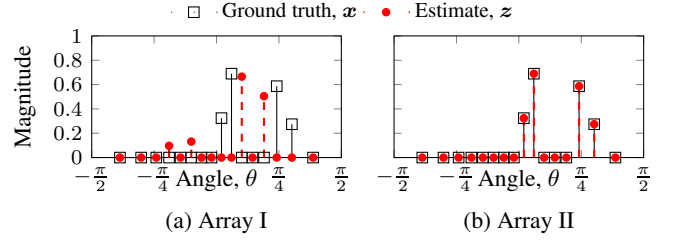\vspace{-0.2cm}
	\section{Conclusions} \label{sec:conclusions}
    This paper investigated the role of the array redundancy pattern in active sensing. We showed that identifying the maximum number of scatterers requires carefully designing the redundancy pattern when employing fewer independent waveforms than transmitters. Specifically, an unfavorable choice of array geometry may fundamentally hamper identifiability such that no waveform will improve identifiability to the level of another array geometry employing the same (reduced) waveform rank, number of physical/virtual sensors, and identical (contiguous) sum co-array. Several open questions for future work emerge from this insight. For example, how severely can identifiability be affected by a poor choice of array or waveform? Furthermore, what impact does a reduced Kruskal rank have in practice when worst-case performance is not of primary interest?

    \bibliographystyle{IEEEtran}
    \bibliography{IEEEabrv,references}
	
\end{document}